\documentclass{aa}
\usepackage{epsfig,times}
\usepackage{color}
\begin{document}
\thesaurus{12         
          (02.01.2;   
           02.13.2;   
           03.13.4;   
           08.13.1;   
           09.10.1;   
           11.10.1) } 
\title{Collimating, relativistic, magnetic jets from rotating disks}
\subtitle{The axisymmetric field structure of relativistic jets 
and the example of the M87 jet}
 
\author{Christian Fendt \and Elisabetta Memola}

\institute{
Astrophysikalisches Institut Potsdam, An der Sternwarte 16,
D-14482 Potsdam, Germany; email: cfendt@aip.de ; ememola@aip.de
}

\date{Received ??; accepted ??}

\authorrunning{Fendt \& Memola}
\titlerunning{Relativistic magnetic jets}

\maketitle

\def\omf{\Omega_{\rm F}}
\def\omk{\Omega_{\rm K}}
\def\rl{R_{\rm L}}
\def\ro{R_{\rm 0}}
\def\rj{R_{\rm jet}}
\def\rf{R_{\rm D}}
\def\ra{R_{\rm A}}
\def\rs{R_{\rm S}}
\def\\psid{\Psi_{\rm D}}
\def\xj{x_{\rm jet}}
\def\xl{x_{\rm L}}
\def\gi{g_{\rm I}}
\def\tx{\tilde{x}}
\def\xaj{x_{A,\Psi = 1}}
\def\xdj{x_{D,\Psi = 1}}

\def\bp{B_{\rm P}}
\def\bh{B_{\phi}}

\def\cm{{\rm cm}}
\def\msun{{\rm M}_{\sun}}
\def\rsun{{\rm R}_{\sun}}

\def\ln{{\rm ln}}
\def\tan{{\rm tan}}
\def\Reo{R_{\Omega}}

\begin{abstract}
We investigate the axisymmetric structure of collimating, relativistic,
strongly magnetized (force-free) jets.
In particular, we include the differential rotation of the foot
points of the field lines in our treatment.
The magnetic flux distribution is determined by the solution
of the Grad-Shafranov equation and the regularity condition
along the light surface.
With differential rotation, 
i.e. the variation of the iso-rotation parameter $\omf $, the 
shape of the light surface is not known a priori and must be
calculated in an iterative way.

For the first time, we have calculated the force-free magnetic 
structure of truly two-dimensional, relativistic jets,
anchored in a differentially rotating disk.
Such an approach allows for a direct connection between
parameters of the central source (mass, rotation) and the 
extension of the radio jet.
In particular, this can provide a direct scaling of the location
of the asymptotic jet light cylinder in terms of the central 
mass and the accretion disk magnetic flux distribution.

We demonstrate that differentially rotating jets must be collimated to a
smaller radius in terms of the light cylinder if compared to jets with  
rigid rotation. Also, the opening angle is smaller.
Further we present an analytical estimate for the jet opening angle
along the asymptotic branches of the light surface.
In general, differential rotation of the iso-rotation parameter leads
to an increase of the jet opening angle.

Our results are applicable for highly magnetized, highly collimated,
relativistic jets from active galactic nuclei and Galactic superluminal
jet sources.
Comparison to the M87 jet shows agreement in the collimation distance.
We derive a light cylinder radius of the M87 jet of 50 Schwarzschild
radii.

\keywords{ Accretion, accretion disks --
           MHD --
           Methods: numerical --
           ISM: jets and outflows --
           Galaxies: individual: M87 --
	   Galaxies: jets }

\end{abstract}

\section{Formation of magnetic jets}
%
Observations of astrophysical jet sources have now established the idea
that jet formation is always connected to the presence of an accretion
disk and strong magnetic fields.
This holds for various scales of energy output, jet velocity and nature
of the jet emitting objects.
Examples are jets from active galactic nuclei (AGN),
Galactic superluminal jet sources,
the example of a mildly relativistic jet from a neutron star (SS 433)
and the numerous class of protostellar jets
(see Zensus et al. 1995; Mirabel \& Rodriguez 1995; Mundt et al. 1990;
Ray et al. 1996).
Magnetic jets are believed to originate very close to the central object
in the interaction region with the accretion disk.
Beside observational arguments also theoretical considerations have shown
that magnetic fields play an important role in jet formation
and propagation 
(Blandford \& Payne 1982; Pudritz \& Norman 1983; 
Shibata \& Uchida 1985; Sakurai 1985; Camenzind 1987;
Lovelace et al. 1991).

If the central object is a black hole as it is the case for AGN and
Galactic superluminal jet sources, the surrounding accretion disk is
the only possible location for a field generation (by dynamo action
or/and advection of flux).
In the case of stellar objects (protostars, white dwarfs or neutron
stars), the central star also carries a relatively strong magnetic
field and it is not yet clear, whether the jet magnetosphere originates
in the disk or in the star.
However, a strong interaction between stellar field and accretion flow
is evident.
The jet formation process itself is not yet fully understood theoretically.
In particular, for the mass transfer from the disk into the jet and the 
process of magnetic field generation a complete physical model is missing.

However, over the last decades the basic ideas of Blandford \& Payne (1982)
have been extended by various authors. 
The general picture is the following.
Matter is lifted from the disk into the magnetosphere and becomes magnetically
accelerated (Ferreira 1997).
Toroidal magnetic fields, generated by inertial back-reaction of the plasma
on the poloidal field, may collimate the disk magnetosphere
into a highly collimated jet flow 
(Camenzind 1987; Chiueh et al. 1991; Lovelace et al. 1991).
In general, due to the complexity of the MHD equations, stationary
relativistic models of magnetic jets has to rely on simplifying 
assumptions such as self-similarity (Contopoulos 1994, 1995), some other
prescription of the field structure 
(Li 1993, Beskin 1997) or the restriction to
asymptotic regimes (Chiueh et al. 1991; Appl \& Camenzind 1993; 
Nitta 1994, 1995).
For highly magnetized jets the force-free limit applies.
This allows for a truly two-dimensional calculation of the
magnetic field structure
(Fendt et al. 1995; Fendt 1997a).
Relativistic magnetohydrodynamics implies that poloidal electric fields,
which are not present in Newtonian MHD, cannot be neglected anymore.

From the observations we know that extragalactic jets as well as
Galactic superluminal jets and protostellar jets are collimated almost 
to a cylindrical shape
(Zensus et al. 1995; Ray et al. 1996; Mirabel \& Rodriguez 1995).
Theoretically, it has been shown that current carrying relativistic jets
must collimate to a cylinder (Chiueh et al. 1991).
For the asymptotic limit of a cylindrically collimated magnetic
relativistic
jet, Appl \& Camenzind (1993) found a non-linear analytical solution 
for the trans-field force-balance in the case of a constant iso-rotation
parameter.
These results were further generalized for jets with differential
rotation (Fendt 1997b).
Such an asymptotic field distribution is especially interesting for jets
emerging from (differentially rotating) accretion disks.

In previous papers, we applied the asymptotic jet model of 
Appl \& Camenzind (1993) as a boundary condition for the calculation 
of {\em global, two-dimensional,} force-free jet magnetospheres
for rapidly rotating stars (Fendt et al. 1995)
or rapidly rotating black holes (Fendt 1997a).
In this paper, we continue our work on 2D force-free jet
magnetospheres applying an asymptotic jet with 
{\em differential rotation}
of the iso-rotation parameter $\omf$ as boundary condition for
the global jet structure.
Such an approach allows for a connection between the differentially
rotating
jet source -- the accretion disk -- and the asymptotic collimated jet.
Since jet motion seems intrinsically connected to the accretion disk,
differential rotation of the field lines should be a natural ingredient
for any magnetic jet structure.
As a principal problem for differentially rotating relativistic jet
magnetospheres,
position and shape of the singular light surface are not known
{\it a priori}, 
but have to be calculated in an iterative way together with the 
magnetic flux distribution.

In Sect.\,2 we recall some basic equations of the theory of relativistic
magnetospheres and discuss several difficulties with the solution of the 
Grad--Shafranov (hereafter GS) equation. 
After some comments on the numerical approach in Sect.\,3,
we discuss our results in Sect.\,4.
A summary is given in Sect.\,5. 

\section{Structure of magnetic jets}

Throughout the paper we apply the following basic assumptions:
{\em axisymmetry}, {\em stationarity} and {\em ideal MHD}.
We use cylindrical coordinates $(R,\phi,Z)$ or $(x,\phi,z)$ if
normalized.
The term {\em 'asymptotic'} always denotes the limit of 
$Z >> R$ unless explicitly stated otherwise.
We consider jets with a {\em finite} radius, $R < \infty$
for $Z\rightarrow \infty$.

\subsection{The force-free, cross--field force--balance}

With the assumption of axisymmetry, a magnetic flux function can
be defined
\begin{equation} 
\Psi = \frac {1}{2 \pi} \int {\vec {B}}_{\rm P} \cdot d{\vec{A}} ,\quad\quad
 R{\vec{B}}_{\rm P} = \nabla \Psi \wedge  {\vec{e}}_{\phi },
\end{equation}
measuring the magnetic flux through a surface element with radius $R$ 
and, in a similar way, the poloidal current through the same area
\begin{equation} 
I = \int {\vec{j}}_{\rm P} \cdot d{\vec{A}} = -\frac{c}{2}\,RB_{\phi},
\end{equation}
which, in the force-free case, flows parallel to the flux surfaces, $I = I(\Psi)$.

The structure of the magnetic flux surfaces is determined by the toroidal
component of Amp\'ere's law, $\nabla \times \bp = 4\pi j_{\phi} /c $, where 
the toroidal electric current density has to be calculated from the
equation of motion projected perpendicular to the flux surfaces
(Camenzind 1987; Fendt et al. 1995).
For strong magnetic fields, inertial forces of the matter can be neglected.
This is the {\em force-free} limit and the equation of motion reduces to
$0=c \rho_c\vec{E} + \vec{j}\times \vec{B}$ with the charge density $\rho_c$.

Combining Amp\'ere's law and the force-free equation of motion the cross-field
force-balance can be written as the modified relativistic GS equation,
\begin{eqnarray} 
R\nabla\cdot\left({\frac{1-(R\omf(\Psi)/c)^2}{R^2}}\nabla\Psi\right) =
& - & \frac{4}{c^2}\frac{1}{R}\frac{1}{2}\left(I^2(\Psi)\right)'\\
& - & R\,|\nabla\Psi|^2 \frac{1}{2}\left(\omf^2(\Psi)\right)',
\nonumber 
\end{eqnarray}
where the primes denote the derivative $\frac{d}{d\Psi}$
(see Camenzind 1987; Okamoto 1992).

$\omf$ is conserved along the flux surfaces, $\omf = \omf(\Psi)$.
We will call it the {\em iso-rotation} parameter, defined by Ferraro's
law of iso-rotation.
It can be understood as the angular velocity of the plasma, reduced
by the slide along the field lines.
Sometimes, it is called the angular velocity of the field lines.
Both, the current distribution $I(\Psi)$ and the rotation law $\omf(\Psi)$
determine the source term for the GS equation and govern
the structure of the magnetosphere.
We apply the following normalization,
\begin{eqnarray} 
R, Z & \Leftrightarrow & x\,\ro , z\,\ro,\nonumber \\
\omf & \Leftrightarrow & \omf\,(c/\ro)\,, \nonumber \\
\Psi & \Leftrightarrow & \Psi\,{\Psi }_{\rm max}\,,\nonumber\\
I & \Leftrightarrow & I\,I_{\rm max}\,. \nonumber
\end{eqnarray}
As the length scale for the GS equation (3) the asymptotic radius
$\ro$ of the light surface is selected (see below).
In order to allow for a direct comparison to rigidly rotating magnetospheres,
the normalization was chosen such that $\omf = 1$ at $ x = 1 $.
With the chosen normalization, Eq.\,(3) can be written dimensionless,
\begin{eqnarray} 
x\nabla\cdot\left({\frac{1-x^2\omf^2(\Psi)}{x^2}}\nabla\Psi\right) =
& - & \frac{1}{x}\frac{g}{2}\left(I^2(\Psi)\right)'
\nonumber \\
& - & x|\nabla\Psi|^2\frac{1}{2}\left(\omf^2(\Psi)\right)'.
\end{eqnarray}
$g$ is a coupling constant describing the strength of the current term
in the GS equation,
\begin{displaymath}
g = \frac {4 I_{\rm max}^2 \ro^2}{c^2 {\Psi }_{\rm max}^2} = 4\,
\left(\frac{I_{\rm max}}{10^{18} {\rm A}}\right)^{\!\!2} 
\left(\frac{\ro}{10^{16} {\rm cm}}\right)^{\!\!2}
\left(\frac{{\Psi }_{\rm max}}{10^{33}\,{\rm Gcm}^2}\right)^{\!\!-2} 
\end{displaymath}
where the parameters are chosen for extragalactic jets.
In this paper, $g$ is in accordance with the definition in Fendt et 
al. (1995) and differs from the definition in Appl \& Camenzind (1993) 
by a factor of two, $g_{\rm Fendt} = 2\,g_{\rm AC}$\footnote{Due to
the fact that the jet radius (where $\Psi=1$) is not known before the
asymptotic GS equation has been solved (Fendt 1997b),
the normalization with $g$ leads to a current distribution $I(\Psi)$
which is {\em not} normalized to unity. 
This difference in normalization is ``hidden'' in the coupling
constant $g$, which could, in principal, be re-scaled appropriately.}.
Interestingly, a coupling constant, defined in a similar way also for
the differential rotation term, would be equal to unity.
The GS equation is numerically solved applying the method of finite
elements (see Appendix).

Along the light surface, where $D\equiv 1-x^2\omf^2(\Psi)=0$,
the GS equation 
reduces to the regularity condition,
\begin{equation}
\nabla\Psi\cdot\nabla D = 
-g\frac{1}{2}\left(I(\Psi)^2\right)' -
\frac{1}{2} |\nabla\Psi|^2\left(\ln\left(\omf(\Psi)^2\right)\right)', 
\end{equation}
which is equivalent to a Neumann boundary condition.
However, for differentially rotating magnetospheres with 
$\omf = \omf(\Psi)$
the shape of this surface is not known a priori and has to be calculated 
in an iterative way together with the two-dimensional solution of the GS 
equation.
For constant $\omf$ the light surface is of cylindrical shape.
As we have shown in a previous publication (Fendt et al. 1995), 
our finite element code satisfies the regularity condition
{\em automatically}, since the surface integral along the light
surface vanishes.

\subsection{Discussion of the force-free assumption}
%
It is clear that relativistic jets must be highly magnetized.
Only a high plasma magnetization gives jet velocities close
to the speed of light (Fendt \& Camenzind 1996).
Therefore, for the calculation of {\em field structure} the 
force-free limit seems to be reasonable.
However, one may question the assumption of a force-free 
{\em asymptotic} jet.
In a fully self-consistent picture of magnetic jet formation, the
asymptotic jet is located beyond the collimating, non force-free
wind region and beyond the fast magnetosonic surface.
The asymptotic jet parameters are determined by the wind motion.
Thus, poloidal current and iso-rotation parameter of the field 
are {\em not} functions free of choice.
The force-free region of a jet is located close to its origin, where
the speed is low.
Beyond the Alfv\'en surface plasma kinetic energy dominates the
magnetic energy, which is just the contrary to force-freeness.

For small plasma density, the Alfv\'en surface of the wind flow
approaches the light surface.
In this case the fast magnetosonic surface moves to infinity 
for a conical flow.
Okamoto (1999) argues that a force-free field distribution extending
to infinity in both $x$ and $z$ direction will asymptotically be
of conical shape, i.e. un-collimated.
However, his approach differs from ours in the sense that he 
{\em assumes} that all field lines will cross the light cylinder. 
Such an assumption {\it per se} prohibits any collimation.
On the other hand, perfect jet collimation is an observational fact.
Astrophysical jets (of very different energy scales)
appear collimated to cylinders of finite radius.

In general, the non force-free relativistic GS equation shows 
three inertial contributions,
\begin{eqnarray}
 0 & = & - \tilde{\kappa} \left(1 - M^2 - x^2\omf^2\right) 
+ \left(1 - x^2\omf^2\right) \frac{\nabla_{\!\!\perp}\bp^2}{8\pi}
+ \frac{\nabla_{\!\!\perp}B_{\phi}^2}{8\pi} \nonumber \\
 & + & \nabla_{\!\!\perp}P 
+ \left(\frac{B_{\phi}^2}{4\pi} -\rho u_{\phi}^2\right)
\frac{\nabla_{\!\!\perp}x}{x}
- \frac{\bp^2\omf}{4\pi}\nabla_{\!\!\perp}(x^2\omf),\nonumber 
\end{eqnarray}
where $\nabla_{\!\!\perp}$ indicates the gradient perpendicular to
$\Psi$,
$\tilde{\kappa} \equiv \kappa\bp^2/4\pi =
 {\bf n}\cdot ({\bf B_p}\cdot\nabla){\bf B_p}/4\pi $
the poloidal field curvature,
$\rho$ the mass density, $u_{\phi}$ the toroidal velocity,
$P$ the gas pressure
and $M$ the Alfv\'en Mach number
(Chiueh et al. 1991).
One can show that in the asymptotic, cylindrical jet the contribution
of inertial terms in the force-balance across the field is weak.
The contribution of gas pressure is usually negligible in astrophysical
jets.
Also, the centrifugal force does not play a role for radii larger than
the Alfv\'en radius, 
since outside the Alfv\'en surface (where $M^2=1-x^2\omf^2$) the plasma
moves with constant angular momentum.
The curvature term vanishes in cylindrical geometry.
Therefore, since for cylindrical jets the contribution from inertial
terms is weak, the configuration is comparable to the force-free case.
The source term of the GS equation may be reduced to a form similar to the
common force-free limit.
We suggest the phrase ``quasi force-free'' for such a configuration
because the GS equation looks force-free even if
the physical system is not magnetically dominated.

In the force-free limit of a highly magnetized plasma the previous
arguments also apply.
However, in difference to the asymptotic regime considered above,
the low plasma density implies that inertial terms are
weak over the {\em whole} two-dimensional jet region.
The centrifugal term $\rho u_{\phi}$ is weak even if the Alfv\'en
surface now comes close to the light surface.
Numerical calculations of the plasma motion along the field have shown
that, for a high magnetization, the Alfv\'en Mach number $M$ grows almost
linearly with radius but remains relatively low (Fendt \& Camenzind 1996).
Thus, the inertial curvature term should not play a dominant role.
Contopoulos \& Lovelace (1994) find from self-similar solutions
that centrifugal forces are dominated by magnetic forces leading
to a re-collimation of the outflow.

In summary, our discussion of the inertial terms in the force-balance
equation has shown
that these terms are generally weak in the case of a high magnetization.
We therefore think that for the calculation of the magnetic field structure
in relativistic jets the force-free assumption is acceptable.
The main motivation of the force-free assumption is clearly the reason of
simplification.
There is yet no other way to calculate a truly two-dimensional field
distribution for relativistic jets.
Naturally, with a force-free solution, nothing can be said about the
flow acceleration itself.

\subsection{Location of the asymptotic light cylinder}
%
The radius of the asymptotic light cylinder $\ro $ is the natural 
length-scale for the GS solution.
A scaling of the GS solution in terms of the properties of the central
object (e.g. its mass) relies on the proper connection between the
asymptotic jet and the accretion disk.
This is feasible only if differential rotation $\omf(\Psi)$ is included
in the treatment (see Sect.\,3).

In the following we consider the location of the light surface and
its relation to the relativistic character of the magnetosphere from
a general point of view.
The light cylinder of a flux surface $\Psi$ is defined as
a cylinder with radius $R =\rl(\Psi) \equiv c/\omf(\Psi)$.
At this position the GS equation becomes singular.
However, this light cylinder is only important for the field line
{\em if} the field line actually intersects it
as for $\Psi_{\rm out}$ in Fig.\,1.
Only then, relativistic effects become dominant.
For example, the poloidal electric field scales with the radius in units
of the light cylinder radius,
$E_{\rm P} = (R/\rl) B_{\rm P}$.
On the other hand, in the case of $\Psi_{\rm in}$ in Fig.\,1, the 
asymptotic radius of the flux surface is smaller than its
light cylinder radius $\rl(\Psi_{\rm in})$ 
(located between $\Psi_{\rm in}$ and $\Psi_0$), 
therefore relativistic effects are small.
For jet solutions with rigid rotation $\omf$ all flux surfaces have
the same light cylinder radius.
Thus, the singular light surface of the magnetosphere is a cylinder.
For jet solutions with differential rotation $\omf$ the flux surfaces
have different light cylinder radii.
The singular surface of the magnetosphere is not a cylinder anymore.

It is now interesting to note that the case of differential rotation 
$\omf(\Psi)$ allows for a hypothetical field distribution 
where 
(i) the light radius of most of the flux surfaces is located within the 
jet radius,
but where also
(ii) the asymptotic radius of the flux surfaces is always smaller
than their light radius.
Such a field distribution would not have a singular light surface
and could be considered as ``non relativistic'',
even if the hypothetical light radii of many field lines are inside
the jet radius.
Such a situation is not possible if the magnetosphere is constrained 
by a constant rotation $\omf$.
This underlines the importance of the treatment of differential 
rotation for jets from accretion disks.

A relativistic description for the jet magnetosphere is always required
if the jet contains a flux surface for which the light radius is
smaller than the asymptotic radius.
 
\setlength{\unitlength}{1mm}
\begin{figure}
\parbox{90mm}
\thicklines
\epsfysize=80mm
\put(10,0){\epsffile{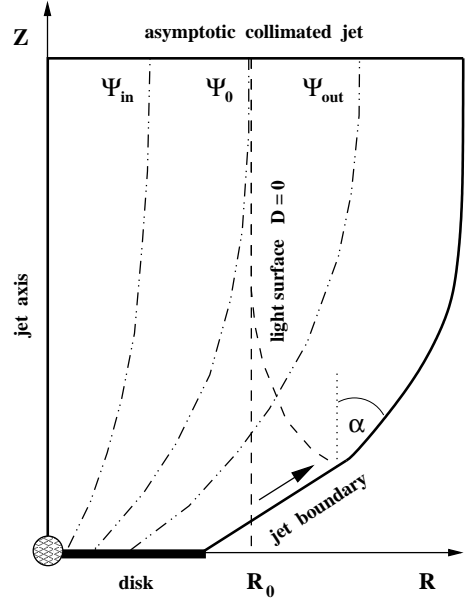}}
\vspace{-0.0cm}
\caption
{Sketch of the jet model.
Axisymmetric jet magnetic flux surfaces $\Psi$ projected into
the meridional plane.
The central object, located within the inner boundary (solid disk),
is surrounded by an accretion disk.
Helical magnetic field lines (laying on the flux surfaces) 
are anchored in the differentially rotating disk 
at the foot points $\rf(\Psi)$.
The jet boundary is defined by the flux surface $\Psi=1$.
The upper boundary condition is a cylindrically collimated jet
solution (Fendt 1997b).
The arrow indicates the numerical deformation of the initially
vertical boundary of the inner solution (at $x=1$) into the 
curved light surface.
The flux surfaces $\Psi_{\rm in}$ ($\Psi_{\rm out}$) have an 
asymptotic radius smaller (larger) than the asymptotic light 
cylinder $\ro$, which is the asymptotic branch of the 
light surface $\rl(\Psi)$ for large $z$.
The flux surface $\Psi_0$ coincides with the light surface 
asymptotically.
The jet half opening angle is $\alpha$ (see Sect.\,2.4, Fig.\,2).
}
\end{figure}

\subsection{The regularity condition and the jet opening angle}

\setlength{\unitlength}{1mm}
\begin{figure}
\parbox{90mm}
\thicklines
\epsfysize=60mm
\put(0,0){\epsffile{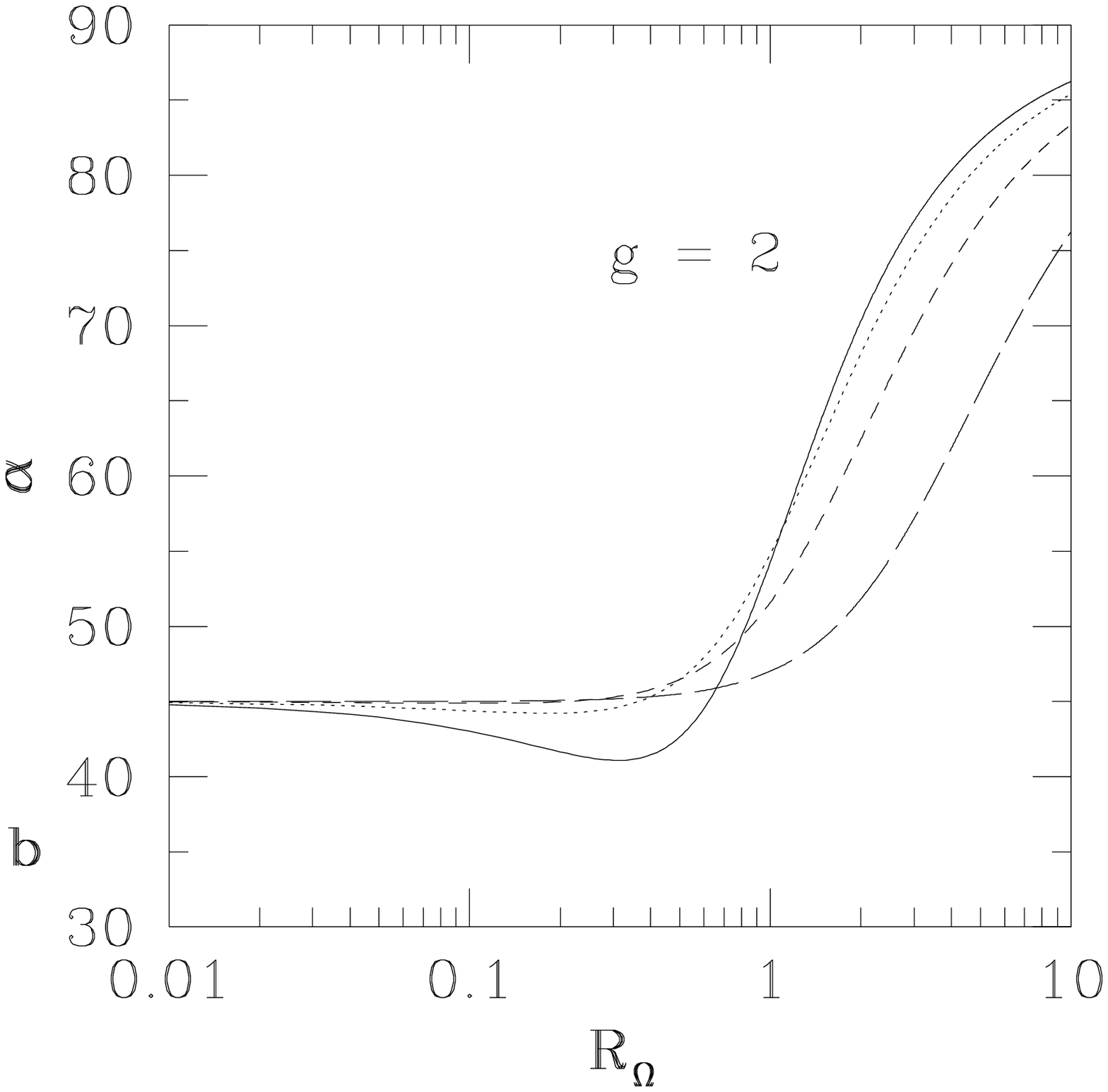}}
\put(0,50){\epsffile{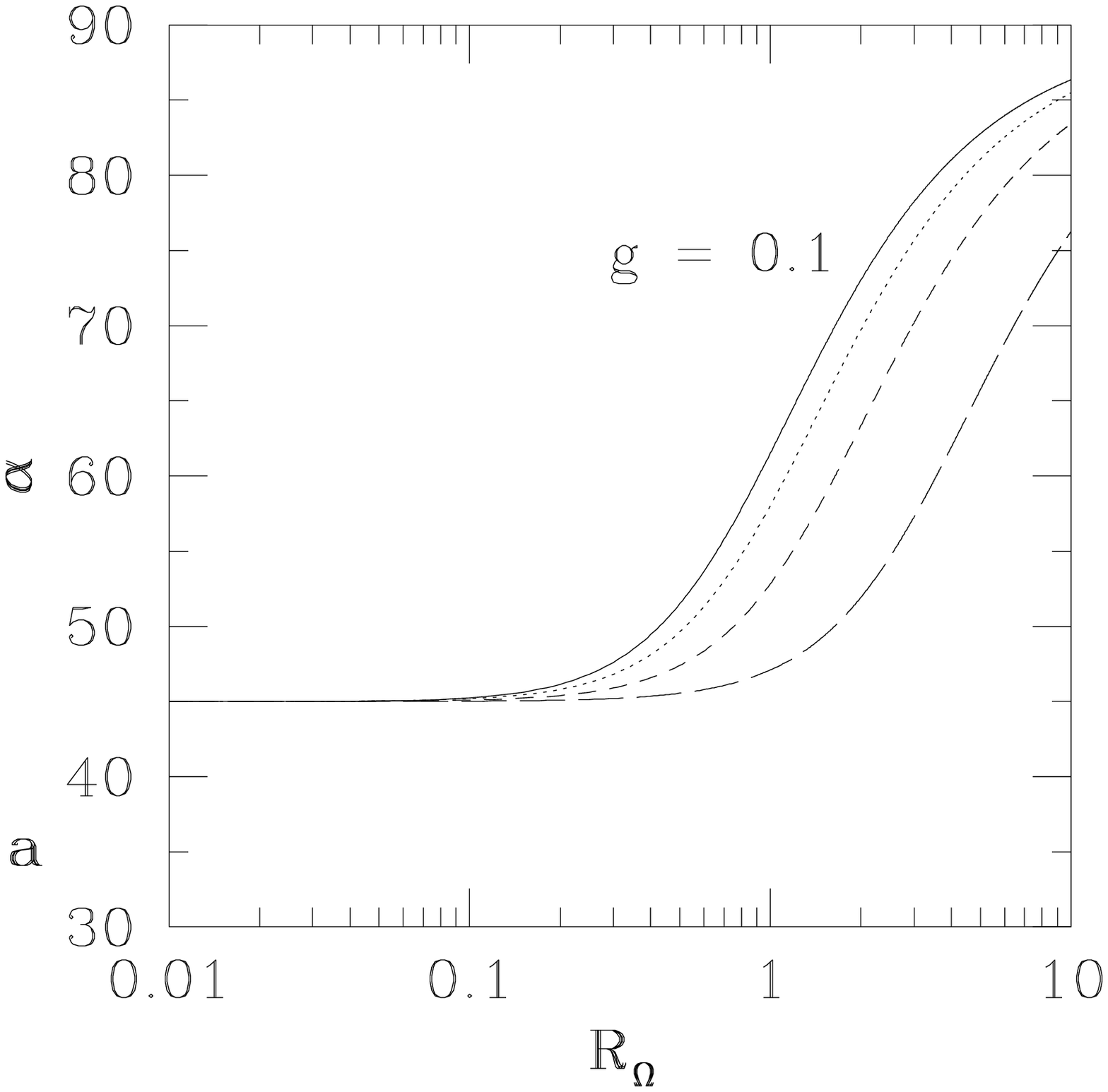}}
\caption
{Jet half opening angle $\alpha(\Psi=1)$ for the analytical 
asymptotic jet solution and along the asymptotic branch of the
light surface in $x$-direction (see Eq.\,(8)). 
Coupling constant $g = 0.1$ ({\bf a}) and $g= 2$ ({\bf b}). 
Asymptotic magnetic flux distribution parameter (Eq.\,(9))
$b =0.5$ (solid), $b=1$ (dotted), 
$b=2$ (short-dashed), $b=5$ (long-dashed).
}
\end{figure}

The regularity condition (5) is the natural boundary condition along
the light surface.
Although it is impossible to solve equation (5) explicitly,
a general relation concerning the jet opening angle can be derived.
First, we rewrite Eq.\,(5) as
\begin{equation}
B_z = \frac{1}{4} g (I^2)'
    - \frac{1}{4} \bp^2\left(\frac{1}{\omf^2}\right)',
\end{equation}
where $\omf(\Psi) = 1 / \xl(\Psi) \equiv \ro/\rl(\Psi)$ has been 
applied.
From Eq.(6) it follows for the radial field component 
$B_x^2 = -g (I^2(\Psi))'/(1/\omf^2(\Psi))'$,
if $\Psi$ intersects the light surface with vanishing $B_z$.
On the other hand, considering a field line perpendicular to the
light surface,
$\nabla\Psi\perp\nabla D$, this provides a condition for the
axial field component,
\begin{equation}
B_z = \frac{g}{2} (I^2)' = \frac{\bp^2}{2}\left(\frac{1}{\omf^2}\right)'.
\end{equation}
Interestingly, this is 
{\em either} only a function of the current distribution $I(\Psi)$
{\em or} depends only from the specification of the rotation
law $\omf(\Psi)$.
Further, in this case it is always $B_z > 0$, since
$(1/\omf^2)' = (\xl^2)' > 0$.
In particular, for the asymptotic ($z\rightarrow\infty$) part of the 
magnetosphere, this implies
that {\em only collimating field lines can cross the light surface}.

Now we consider the asymptotic branches of the light surface.
For the asymptotic branch in $z$-direction it holds 
$(\nabla D)_x >> (\nabla D)_z \simeq 0$.
Further, it is $B_x (\ln \omf^2)' = 0$,
implying either a collimated field structure, $B_x \equiv 0$
or rigid rotation, $(\omf(\Psi))' \equiv 0$.
From this we conclude that in the asymptotic regime of a cylindrical
light surface, also the flux surfaces along this light cylinder
must be of cylindrical shape. 
Collimation must occur in the non-asymptotic region of the jet.

If we now assume that there exists an asymptotic part of the light 
surface in $x$-direction (where $x >> z$) 
and that $(\nabla D)_z >> (\nabla D)_x \simeq 0$, 
we derive an equation for the half jet {\em opening angle},
\begin{equation}
\alpha(\Psi) = \tan^{-1}\left(\sqrt{1 +
 \frac{1}{4}g\frac{(I^2(\Psi))'(\omf^2(\Psi))'}{\omf^4(\Psi)}}\right),
\end{equation}
for the flux surfaces in this region. 
As a general example we apply the analytical solution obtained for the
asymptotic jet (Fendt 1997b),
\begin{eqnarray}
\Psi(x) & \equiv & \frac{1}{b} \ln\left(1+\left(\frac{x}{a}\right)^2\right),
\quad b \equiv \ln\left(1+\left(\frac{x_{\rm jet}}{a}\right)^2\right), 
\nonumber \\
\omf^2(\Psi) & = & 
\frac{g\,b^2}{4}\left(\frac{I^2(\Psi)}{(1-e^{-b\Psi})^2} 
- \frac{1}{(1-e^{-b})^2}\right) + \omf^2(1)
\end{eqnarray}
for Eq.\,(9). Here, $b$ is a measure for the ratio of jet radius to
jet core radius $a$.
Finally, we obtain the half opening angle for the outermost flux 
surface $\Psi =1$,
\begin{equation}
\alpha = 
\tan^{-1}\left(\sqrt{
1 + \Reo \left(4 \Reo\left(\frac{e^b-1}{be^b}\right)^2 
+\frac{g/2}{e^b-1}\right)
}\right),
\end{equation}
where $\Reo$ is defined as $(\omf(1))'/\omf(1)$.
Note, that Eq.\,(10) is only valid in the limit of $(\nabla D)_x \simeq 0$.
Fig.\,2 shows the variation of the opening angle $\alpha$ with the
parameters $a$ and $b$ for two choices of the strength of the poloidal
electric current.
In general, jets with a strong differential rotation $\omf(\Psi)$
(i.e. large $\Reo$) have a larger opening angle.
Also, jets with a large ratio of jet to core radius have a smaller
opening angle.
Therefore, jets originating in a small part of the accretion disk,
equivalent to small value of $\Reo$, will be collimated
to a smaller opening angle.
It is interesting to note that, in the case of a rigid rotation, 
the limiting half opening angle is $45\degr$,
{\em independently} of $g$ and $b$.

\section{The jet-disk connection,  
         providing the true length scale of the GS solution}
%
The natural length scale of the relativistic
GS equation (3) is the asymptotic 
light cylinder $\ro $ (see Sect.\,2.3).
Its size is related to the iso-rotation parameter $\omf(\Psi)$, which 
itself is connected to the angular rotation of the foot points of the
field lines.
Concerning the GS equation, the size of $\ro $ follows purely from
electro-magnetic quantities, if the coupling constant $g$ is chosen.
The GS solution can be scaled to any central object from stars to
galactic nuclei as long as the interrelation of the parameters
${\Psi }_{\rm max}$, $I_{\rm max}$ and $\ro$ provides the same $g$.
So far, no connection has been made to the type of central object.
Here, we treat the question where the asymptotic light cylinder
is located in physical units.

In the case of rigid rotation, the light cylinder radius is usually
estimated by choosing a distinct radial distance from the central
object
and defining $\omf$ under the assumption that the jet magnetosphere
is anchored in that point. 
If the central object is a black hole, the marginally stable orbit 
implies an upper limit for $\omf$.
For jets in AGN this estimate leads to the common conclusion that the
light cylinder radius is at about $10\rs$ 
and the typical jet radius at about $100\,\rl$.
Clearly, such arguments relies on the {\em assumption} that the
field line emerging at this very special radius defining $\omf$
also extends to the light cylinder radius $\rl$ (see Sect.\,2.3).

This picture changes, if differential rotation $\omf(\Psi)$ is
considered.
Different flux surfaces anchor at different foot point radii and
have different light radii (Sect.\,2.3).
Assuming a Keplerian rotation, the light surface radius $\rl(\Psi)$
is located at 
\begin{equation}
\rl(\Psi) = 4\times 10^{15}\cm 
\left(\frac{\rf(\Psi)}{R_{\rm S}}\right)^{\!\!3/2}
\!\!\left(\frac{M}{10^{10}\,\msun} \right)\,,
\end{equation}
where $R_{\rm S}$ is the Schwarzschild radius of a point mass
and $\rf(\Psi)$ the foot point of the flux surface $\Psi$ on a
Keplerian disk.
A more general equation is 
\begin{equation}
\frac{\rl(\Psi)}{R_{\rm S}} = 
\sqrt{2}\left(\frac{\rf(\Psi)}{R_{\rm S}}\right)^{3/2}.
\end{equation}
$\rf(\Psi)$ is determined from the magnetic flux distribution along
the disk and is defined by a certain disk model.
Fig.\,3 shows the location of the light radius $\rl$ for a field line
anchored at a foot point radius $\rf$ in a Keplerian disk
around a central object of mass $M$ (see Eq.\,11). 
Note that the unit of the field line footpoint radius in Eq.\,(12) and
Fig.\,3 is the Schwarzschild radius.
Therefore, Fig.\,3 is appropriate only for relativistic jets.
The footpoint radii for protostellar jets are several {\em stellar}
radii, corresponding to about $10^6$ Schwarzschild radii
(which would be located far above the box in Fig.\,3).

So far, nothing can be said about the location of the asymptotic radius
of the field lines in general.
The essential question is where the asymptotic radius of a flux surface
is located in respect to its light cylinder.
This question can only be answered by a detailed model considering
the {\bf two-dimensional} field distribution
{\em including differential rotation} $\omf(\Psi)$.
Only in such a model, the flux distribution of the asymptotic jet can
be connected to the flux distribution of the ``star''-disk system.
Certainly, both boundary conditions -- asymptotic jet and accretion
disk -- rely on model assumptions.
However, in a self-consistent model these boundary conditions
have to follow certain constraints (see Sect.\,4.1, 4.3). 

\setlength{\unitlength}{1mm}
\begin{figure}
\parbox{80mm}
\thicklines
\epsfysize=70mm
\put(5,0){\epsffile{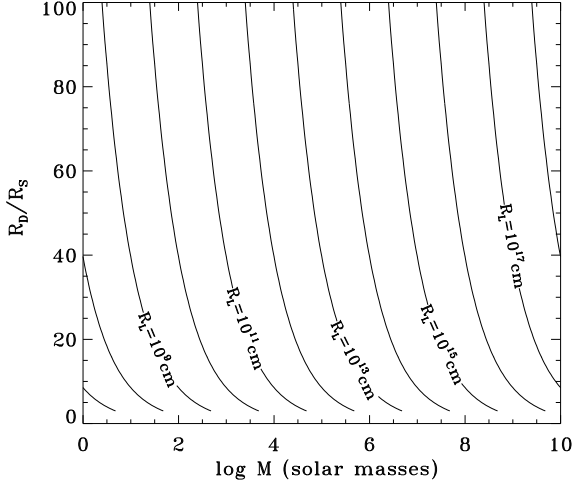}}
\vspace{0.0cm}
\caption
{Location of the light cylinder radius
of a flux surface $\rl(\Psi)$, 
anchored at a certain foot point radius $\rf(\Psi)$
in units of the Schwarzschild radius $R_{\rm S}$
in a Keplerian disk around a point mass $M$.
Note that for non collapsed stellar objects the footpoint radii
of the jet field lines are located at about $10^6 R_{\rm S}$.
}
\end{figure}

\section{The two-dimensional jet solution}
%
\subsection{Disk and jet boundary condition}
%
Three important boundary conditions determine the two-dimensional
flux distribution.
The first boundary condition is in the asymptotic region.
Here we assume a cylindrically collimated jet.
We apply the magnetic flux distribution derived by Fendt (1997b),
where the rigidly rotating jet model of Appl \& Camenzind (1993)
is extended for differential rotation $\omf(\Psi)$.
In particular, our asymptotic jet shows the typical jet core-envelope
structure of magnetic flux and electric current, i.e. a configuration 
where most of the magnetic flux and poloidal electric current is 
concentrated within a ``core'' radius. 
The asymptotic model provides not only the asymptotic magnetic flux
boundary condition but also the $\omf(\Psi)$ and $I(\Psi)$ distribution 
for the whole two-dimensional jet magnetosphere.
In the model of Fendt (1997b) these functions follow from the solution 
of the one-dimensional (asymptotic) GS equation across the cylindrical
jet and the prescription of  
$I(x) = (x/a)^2/(1+(x/a)^2)$ together with 
$\omf^2(x) = e^{h - h x}$,
where $a$ is the core radius of the electric current distribution
and $h$ governs the steepness of the $\omf$-profile\footnote{
For figures of these functions and the related $\Psi(x)$,
$\omf(\Psi)$ and $I(\Psi)$ distribution, we refer to Fendt (1997b).}.

The second boundary condition is the magnetic flux distribution 
along the disk.
This distribution is in general {\em not known} as a solution of the
full MHD disk equations.
%
%
Typical models rely on various simplifying assumptions, as e.g.
stationarity, the distribution of magnetic resistivity or the
disk turbulence governing a dynamo process.
We apply an analytic flux distribution similar to the model of
Khanna \& Camenzind (1992),
who calculated the stationary accretion disk magnetic field structure 
around a super massive black hole.
The typical behavior of the magnetic flux distribution is 
(i) a slight increase of magnetic flux along the innermost disk,
(ii) a small or vanishing flux at the inner disk radius,
(iii) a strong increase of magnetic flux at intermediate radius
(the core radius) and 
(iv) a saturating behavior for large radii.
Using the normalization introduced above, we choose the following
disk boundary magnetic flux distribution,
\begin{equation}
\Psi_{\rm disk}(x) = 
\frac{1}{\tilde{b}}
\ln\left(1 + \left(\frac{x-x_{\rm in}}{\tilde{a}}\right)^2\right)\, 
\end{equation}
with $\tilde{b} = \ln(1 + (x_{\rm disk}-x_{\rm in})^2/\tilde{a}^2))$ 
(see Fig.\,4).
The parameters are: the core radius $\tilde{a}$, the disk outer radius,
$x_{\rm disk}$ and the disk inner radius, $x_{\rm in}$.
For simplicity, we choose $x_{\rm in} \simeq 0$ without loss of generality.
Such a choice will definitely not influence the global jet solution 
which is normalized to the asymptotic light cylinder radius.
%

The third boundary condition is the jet boundary $x_{\rm jet}(z)$.
Along this boundary the flux distribution is $\Psi = 1$ by definition.
However, the shape of this boundary is not known {\it a priori}.
It must be determined by the regularity of the solution across the
light surface (see also Fendt et al. 1995).
A slightly different shape may give the same solution.
However, the main features of the solution as opening angle or 
locus of the collimation are fixed by the internal equilibrium.
Therefore, the regularity condition governs the shape of the jet boundary.
For a given $I(\Psi)$, $\omf(\Psi)$, disk and jet boundary condition,
the jet radius $x_{\rm jet}(z)$ is uniquely determined.

The inner spherical grid boundary with radius $x_{\star}$ close to the
origin, indicates the regime of the central source, possibly enclosing
a collapsed object.
Neutron stars or magnetic white dwarfs may carry their own magnetic
field, 
a black hole can only be thread by the disk magnetic field.
In any case, the magnetic flux distribution is a combination of 
``central'' magnetic flux
and disk magnetic flux $\Psi = \Psi_{\star} + \Psi_{\rm disk}$.
For simplicity we assume that the magnetic flux increases monotonously
from the axis to the disk edge and 
$ \Psi(x_{\star}) = \Psi_{\star}(x_{\star}) $
and $\Psi_{\rm disk}(x_{\star}) = 0$. 

%
\setlength{\unitlength}{1mm}
\begin{figure}
\parbox{90mm}
\thicklines
\epsfysize=52mm
\put(0,0){\epsffile{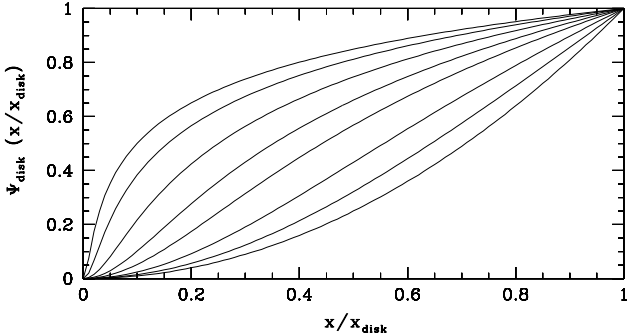}}
\vspace{0.0cm}
\caption
{Magnetic flux distribution along the disk $\Psi_{\rm disk}(x)$ 
as defined in Eq.\,(13).
Parameters: 
$x_{\rm in} = 0$, 
$x_{\rm disk}/\tilde{a} = 100, 40, 15, 7, 4, 2, 1, 0.01$.
}
\end{figure}

\setlength{\unitlength}{1mm}
\begin{figure*}
\parbox{90mm}
\thicklines
\epsfysize=120mm
\put(-10,0){\epsffile{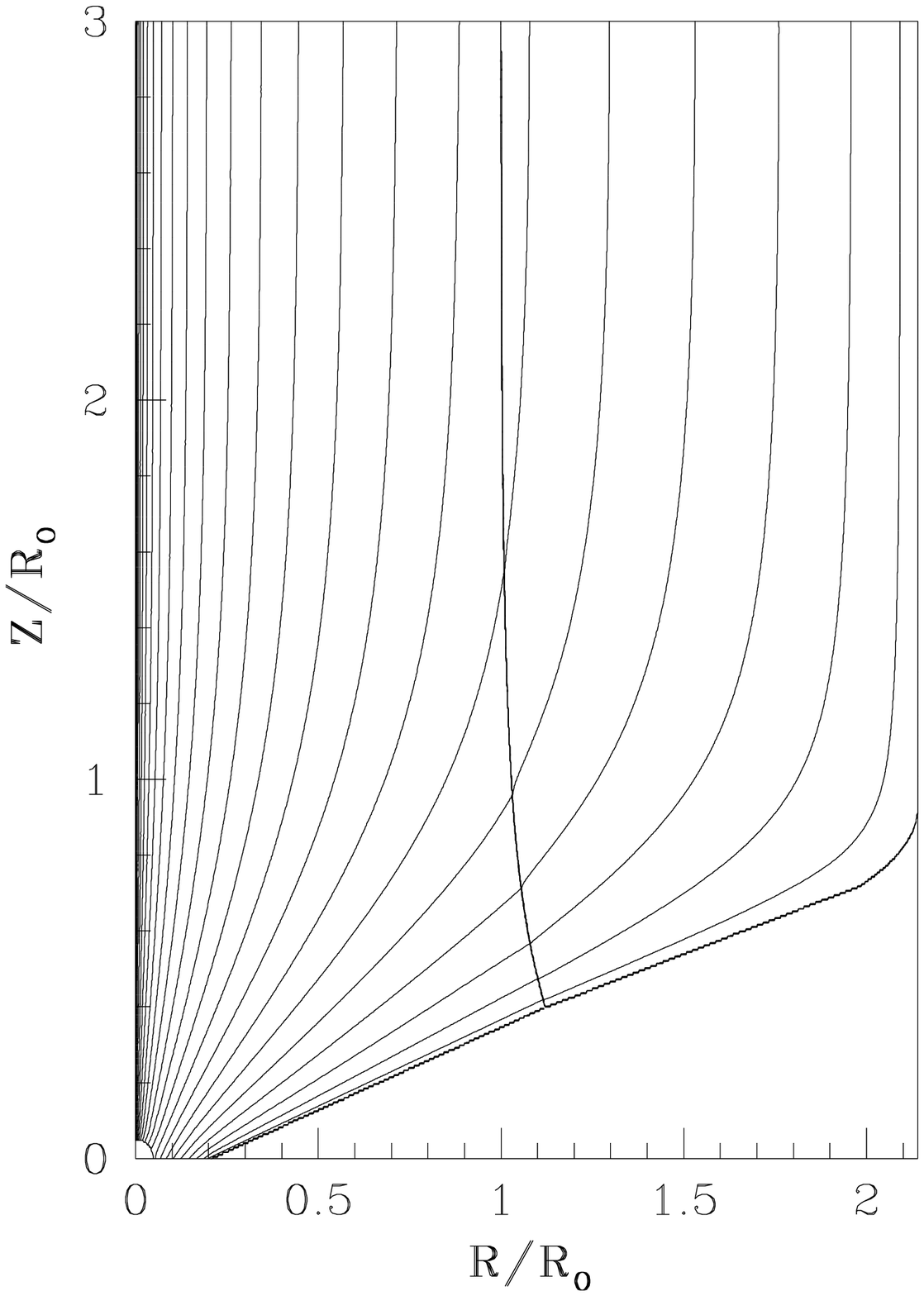}}
\put(80,0){\epsffile{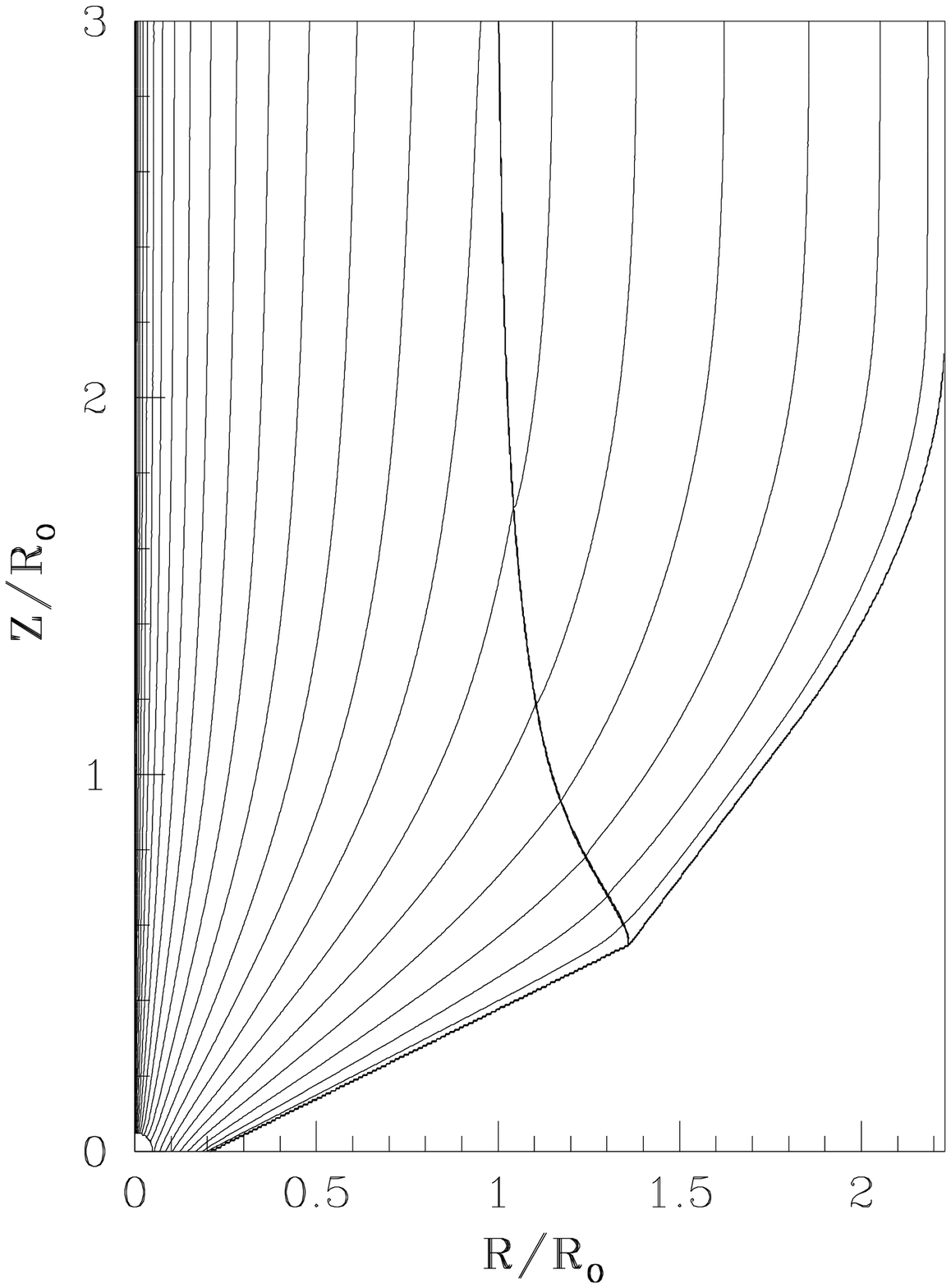}}
\vspace{0.0cm}
\caption
{Two-dimensional magnetic flux distribution $\Psi(x,z)$ for two
different {\em asymptotic} rotation laws. 
{\it Left}: $h=0.2$, $g=2.5$
{\it Right}: $h=0.5$, $g=2.0$.
Shown are iso-contours of the magnetic flux 
(equivalent to poloidal magnetic field lines) with contour levels
$\Psi_n = 10^{-(0.1\,n)^2}$, $n=0,..,25$.
Note that due to the choice of contour levels the iso-contour density
does not mirror the field strength.
}
\end{figure*}

\subsection{The two-dimensional collimating magnetic field structure}
%
Results of numerical solutions of the GS equation are presented in
Fig.\,5.
Shown is the two-dimensional structure of the magnetic flux surfaces
as projection of the helical field lines onto the meridional plane.
In general, for a 
choice of the ``free functions'' $I(\Psi)$ and $\omf(\Psi)$,
here taken from the asymptotic cylindrical jet solution,
the field structure is determined by the boundary
conditions and the regularity condition along the light surface.

We calculated two solutions with a different choice for the steepness
parameter in the iso-rotation $\omf$.
The first solution is for $h=0.2$ (Fig.\,5, {\it left}).
This is more close to the case of rigid rotation.
Indeed the solution look rather similar to the solutions presented
in Fendt et al. (1995).
The second solution is for $h=0.5$ (Fig.\,5, {\it right}).
The steeper profile for the rotation law implies a smaller asymptotic jet 
radius (Fendt 1997b).
This can be seen in comparison with the rigid rotation solutions
(Fendt et al. 1995).
However, a larger poloidal electric current can balance the effect of
differential rotation. 
Therefore, the $h=0.2$ solution (with $g=2.5$) collimates to a smaller
asymptotic jet radius than the $h=0.5$ solution (with $g=2.0$).
A $h=0.2$ solution with $g=2.0$ would have a jet radius of $2.4$.
The second solution with the steeper profile of the rotation law 
$\omf(\Psi)$ 
would better fit to a Keplerian disk rotation.
A perfect match would require an even steeper 
$\omf(\Psi)$-profile (see below).

The mean half opening angle of the jet magnetospheres is about 
$60\degr$.
As discussed above, the shape of the outermost flux surface 
($\Psi = 1$) is {\em not} prescribed but 
is a result of our calculation eventually determined by the
regularity condition.
After crossing the light surface the jets collimate to their
asymptotic radius within a distance from the source of about
$1 - 2\,\ro$ along the jet axis.
The opening angle of the second solution is smaller, however,
the jet collimation is achieved only at a larger distance
from the central source.
In our examples, the ``jet expansion rate'', which we define as the
ratio of the asymptotic jet radius to the foot point jet radius 
(the ``disk radius''), is about 10.
The true scaling of the jet magnetosphere in terms of the size of the
central object can be determined by connecting the jet iso-rotation
parameter $\omf(\Psi)$ to the disk rotation (see next Sect.).

We note that, although in our computations the jet boundary 
$x_{\rm jet}(z)$ is determined by the force-balance within the jet,
and therefore subject to the regularity condition,
with our results we do not prove the magnetohydrodynamic
{\em self-collimation process} of the jet flow.
Clearly, the calculated jet magnetosphere is self-collimated in the
sense that its structure has been determined only by the internal
properties.
However, the actual collimation process of the jet flow from an
un-collimated conical outflow into a cylinder could only be
investigated by time-dependent simulations taking into account
the interaction with the ambient medium.
 
On the other hand, 
we can assume that our jet solution is embedded in an ambient medium.
If we further assume an equilibrium between the internal pressure
(magnetically dominated) and external (gas) pressure 
along the jet boundary,
we may derive the gas pressure distribution in the ambient medium,
since we know the magnetic pressure distribution along the collimating
jet radius.
In this case, the jet solution may be considered as collimated by  
ambient pressure.
 
To our understanding one may claim a self-collimation only,
if the jet flow collimates {\em independently} from external
forces.
Since in our treatment we do not consider the interrelation with
the medium outside the jet, 
we cannot decide whether the flow is self-collimated or pressure
collimated.

The field structure is governed by the choice of the functions
$I(\Psi)$ and $\omf(\Psi)$, here taken from an asymptotic jet solution.
In combination with the disk magnetic flux distribution (13)
we can determine two parameters interesting for the
jet-disk interaction.
These are (i) the magnetic angular momentum loss per unit time and
unit radius from disk into the jet
and (ii) the toroidal magnetic field distribution along the disk.
With $I(\Psi)$ as the angular momentum flux per unit time per unit flux
tube,
the (normalized) angular momentum flux per unit time per unit radius is 
$d\dot{J}/dx = - x B_{\rm z} I(x)$ along the disk.
Fig.\,6 shows the behavior of both quantities for our jet model
with the steeper profile of the rotation law, $h=0.5$.
As we see, most of the magnetic angular momentum is lost in the outer
parts of the disk.
This may have interesting applications for accretion disk models taking
into account a magnetized wind as a boundary condition.
The total magnetic angular momentum loss is determined by the
normalization,
$\dot{J} = - (I_{\rm max} \Psi_{\rm max}/c ) \int I(\Psi) d\Psi $ or
$\dot{J} = - (\sqrt{g}\Psi_{\rm max}/2\ro ) \int I(\Psi) d\Psi $.
The magnetic toroidal field distribution along the disk has a maximum
at about half the disk radius.

Clearly, these parameters are biased by the magnetic flux disk boundary
condition (13) of our model. 
However, we believe that the main features are rather general and valid
for any poloidal current and magnetic flux distribution with the typical
core-envelope structure.

\setlength{\unitlength}{1mm}
\begin{figure}
\parbox{90mm}
\thicklines
\epsfysize=45mm
\put(5,50){\epsffile{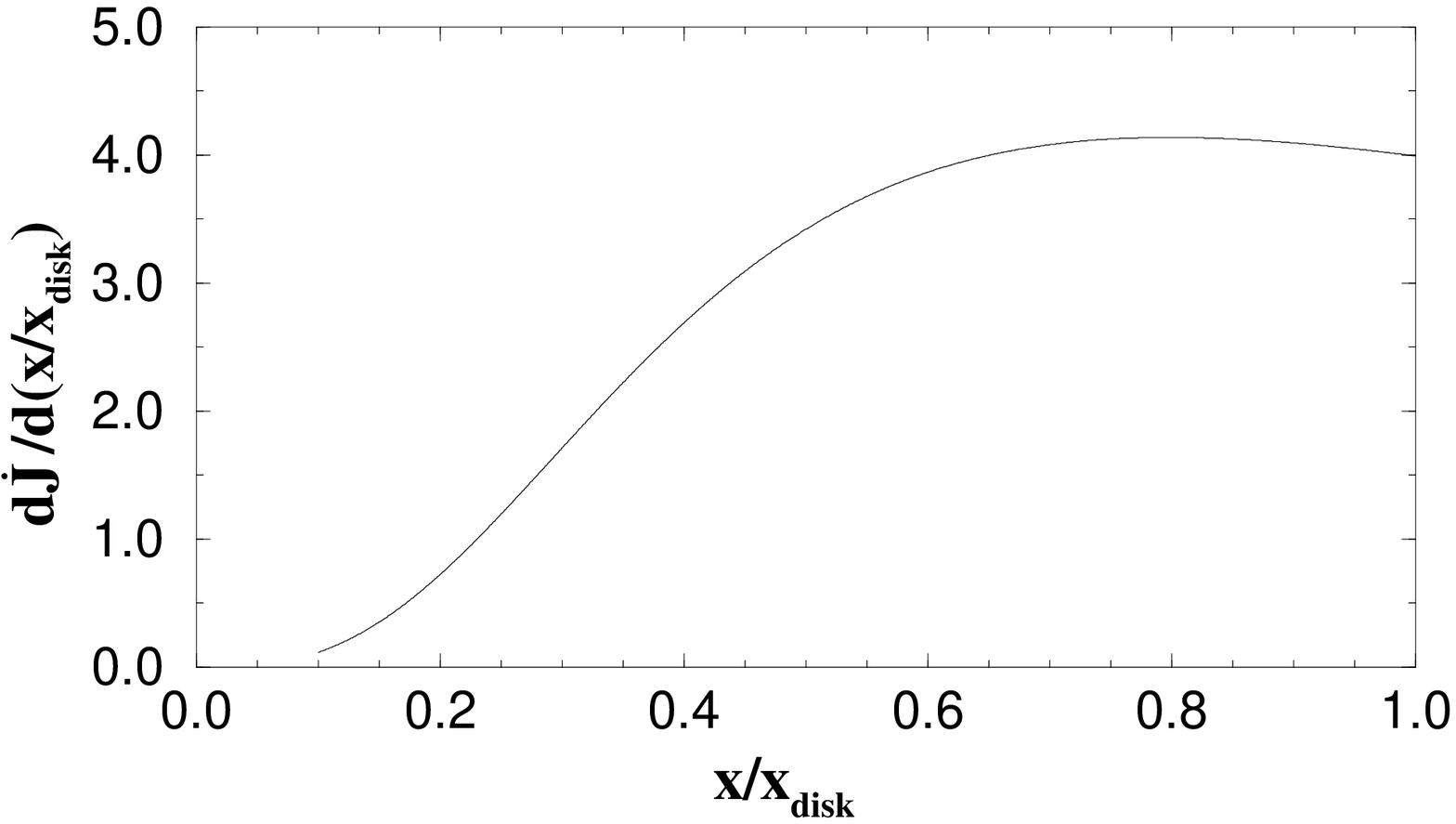}}
\put(5,0){\epsffile{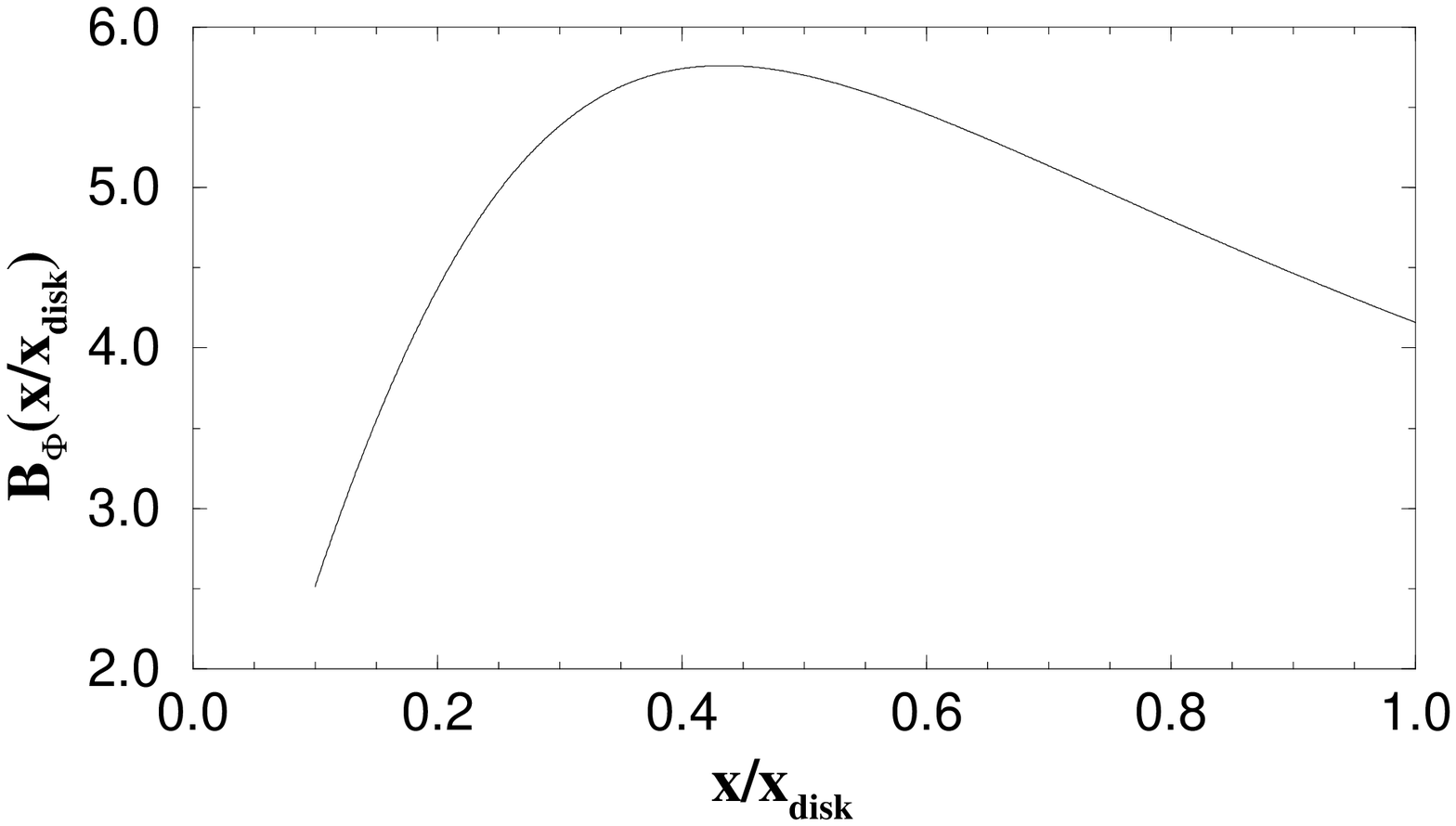}}
\vspace{0.0cm}
\caption
{Magnetic angular momentum loss per time unit per unit radius
$d\dot{J}/dx$ at radius $x$ ({\it above}) and
disk toroidal field distribution $B_{\phi}(x)$ ({\it below}) for
the jet solution with $h=0.5$ shown in Fig.\,5.
}
\end{figure}

\subsection{Scaling relations of disk and jet}
%
As discussed above, the two-dimensional magnetic field distribution
connecting the asymptotic jet region with the lower disk boundary
allows for a direct scaling of the jet in terms of the size of the
central object.
This is simply based on the assumption that the foot points of the field 
lines are rotating with Keplerian speed, $\omf = \omk$
and to the fact that in ideal MHD the iso-rotation parameter $\omf$
is conserved along the field lines.
It is therefore possible to construct a self-consistent model of the
whole ``star''-disk-jet system with only a small set of free parameters.
In the following we will motivate such a model.

The first example demonstrates how the connection between the
asymptotic jet and the disk, applied for our very special model
assumption, provides a specific estimate for the asymptotic 
light cylinder $\ro$.
Normalizing the Keplerian velocity $\omk$ in the same way as $\omf$ 
(Sect.\,2.1), we obtain the expression
\begin{equation}
\ro = \frac{GM}{c^2 \omk^2}\frac{1}{x^3}
    = \frac{GM/c^2}{\omf^2(\Psi=1) x_{\rm disk}^3}
    = \frac{0.5 R_{\rm S}}{\omf^2(\Psi=1) x_{\rm disk}^3}.
\end{equation}
Iso-rotation parameter $\omf(\Psi)$ and disk radius $x_{\rm disk}$
are fixed by our model.
Therefore, the asymptotic light cylinder is proportional to the
mass of the central object.
For $\omf^2(1)=0.54$ (which refers to the $h=0.5$ model)
and $x_{\rm disk}=0.2$ the asymptotic light
cylinder is $\ro = 116 R_{\rm S}$,
which is about 2 times larger compared to the jet solution with
a rigid rotation $\omf \equiv 1$
and will increase for larger values of $h$.
%
%
With the choice of $g$, the value of $\ro$ constraints the maximum
poloidal magnetic flux and electric current.
Here, no assumption is made about the flux distribution along the
disk. 

In the second example we determine the disk magnetic flux distribution
$\Psi(x)$ combining the asymptotic jet rotation law $\omf(\Psi)$ 
with a Keplerian disk rotation $\omk(x)$.
From Eq.\,(14) follows that 
$\omf(\Psi)/\omf(1) = \omk(x)/\omf(\Psi=1)= (x/x_{\rm disk})^{-3/2}$.
In combination with the numerically derived $\omf(\Psi)$ this gives
the $\Psi(x)$ along the disk (Fig.\,7).
The figure shows that the disk flux distribution derived from the
asymptotic jet is distributed only over the outer part of the
disk.
This can be interpreted in two ways. 
First it may imply a relatively large inner disk radius and, hence,
an asymptotic jet radius small in terms of radii of the central object.
Secondly, it just underlines the fact that in our model the distribution
of the asymptotic jet iso-rotation parameter is too flat in order to
be truly connected to a disk magnetic flux with an extended radial 
distribution.
For a model taking into account the disk Keplerian rotation in a 
fully self-consistent way, the magnetic flux distribution which
has to be used as disk boundary condition for the GS solution is the
one derived in Fig.\,7.

On the other hand, the assumption of a Keplerian disk rotation 
together with a certain disk magnetic flux distribution provides an 
expression for the iso-rotation parameter 
$\omf(\Psi) = \omk(x(\Psi)) = 
(GM/\ro c^2) 
\left(
{\tilde{a}}^{2}\left(e^{\tilde{b}\Psi}-1\right)\right)^{-3/2}$.
%
%
Here, the disk magnetic flux distribution (13) has been used.
Eventually, one finds
\begin{equation}
\frac{\omk(\Psi)}{\omf(1)} =
\left(\frac{x_{\rm disk}}{\tilde{a}}\right)^{3/2}
\left(\left(1 + 
\left(\frac{x_{\rm disk}}{\tilde{a}}\right)^2\right)^{\Psi}
-1\right)^{-3/4}.
\end{equation}
This function is definitely steeper compared to the 
$\omf(\Psi)$-distributions which have been derived in Fendt (1997b)
and are used in the present paper.
Here, we see the limitation of our ansatz.
A steeper profile for rotation law is not yet possible to treat with our code
due to the lack of numerical resolution.
The non-linear character of the GS equation becomes more problematic 
due to the gradients in the $\omf$-source term.

In summary, only a model including differential rotation $\omf(\Psi)$
may provide a connection between the asymptotic jet, the disk 
magnetic flux distribution and also the size of the central object.
With our model we have presented a reasonable first solution for
a self-consistent treatment. 

\setlength{\unitlength}{1mm}
\begin{figure}
\parbox{90mm}
\thicklines
\epsfysize=52mm
\put(0,0){\epsffile{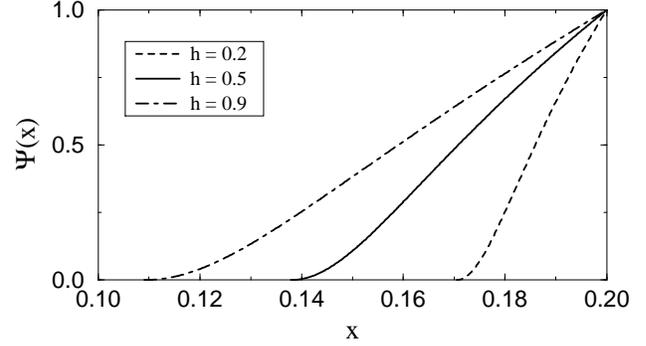}}
\vspace{0.0cm}
\caption
{Magnetic flux distribution $\Psi(x)$ along the disk surface as
determined from the asymptotic jet properties and the Keplerian
rotation of the disk.
}
\end{figure}

\subsection{Application to the M87 jet}
%
The jet of M87 shows superluminal motion clearly indicating
a highly relativistic jet velocity (Biretta et al. 1999).
Recent radio observations have been able to resolve the innermost 
region of the M87 jet formation region 
with $0.33 \times 0.12$ mas beam resolution
(Junor et al. 1999),
corresponding to $2.5 - 7.0 \times10^{16}\cm$.
Assuming a central supermassive black hole of $3\times10^9\msun$ 
(Ford et al. 1994), this is equivalent to about $30\,\rs$!
%
%
The derived jet full opening angle is $60\degr$ up to a
distance of 0.04 pc from the source with a ``strong collimation'' 
occurring afterwards (Junor et al. 1999).

We now apply our two-dimensional jet model to these observations
and compare the geometrical scales.
Such a comparison is not possible for e.g. self-similar models.
From the observed radio profile resolving the inner M87 jet
(see Fig.\,1 in Junor et al. 1999), 
we deduce a jet radius of about 120 Schwarzschild radii. 
With this, the first important conclusion is that the ratio of jet
radius to light cylinder radius must be definitely less than the
value of 100 which is usually assumed in the literature.
A number value of 3-10 seems to be much more likely.
Numerical models of two-dimensional general relativistic magnetic jets
fitting in this picture were calculated by Fendt (1997a).
These solutions, however, do not take into account the differential
rotation $\omf(\Psi)$.

Junor et al. (1999) claim that the M87 jet radius in the region
``where the jet is first formed cannot be larger than'' their
resolution of $30\,\rs$.
Our conclusion is that the expansion rate is limited in both directions.
The new radio observations give a minimum value of 3. 
Theoretical arguments limit the expansion rate to the value of 
about 20, since the jet mass flow must originate outside the marginally
stable orbit which is located at $3-6 \rs$.
Clearly, if the jet radius is really as small as observed in M87,
general relativistic effects may vary the field structure
in the jet formation region.

From our model solutions, we derive a light cylinder radius of the M87
jet of about $50\,\rs$.
The value derived from Eq.\,(14) differs from that by a factor of
two, but is biased by the unknown size of the disk radius
$x_{\rm disk}$.
This parameter, however, does not affect the global solution.
Considering the standard relativistic MHD theory, nothing special is
happening at the light cylinder.
For a highly magnetized plasma wind the light surface corresponds
to the usual Alfv\'en surface which itself does not affect the flow 
of matter.
Hence, the light cylinder is un-observable.

Also the opening angle in our numerical solution is larger than the 
observed value by a factor of two.
This cannot be due to projection effects since any inclination 
between jet axis and the line of sight will increase the observed
opening angle.
We hypothesize that a numerical model with a steeper profile for the 
iso-rotation parameter
will give a smaller jet opening angle comparable to the observed data.
This is not surprising, since the jet footpoint anchored in a Keplerian
disk rotates faster than in our model.
Nevertheless, comparing the collimation distance observed in the M87
jet and assuming a similar ratio of jet radius to light cylinder radius
as in our model with $h=0.5$, we find good agreement. 
The collimation distance is $2\ro$.

In summary, we conclude that the example of the M87 jet gives
clear indication that the light cylinder of AGN jets might not be 
as large as previously thought.
Although our model does not fit the observed geometrical properties
of the inner M87 jet perfectly,
we find in general a close compatibility.

\section{Conclusions}
%
We have investigated the two-dimensional magnetic field distribution in
collimating, relativistic jets.
The structure of the axisymmetric magnetic flux surfaces is calculated 
by solving the relativistic force-free Grad-Shafranov equation 
numerically.
In relativistic MHD, {\em electric fields} become important in
difference to Newtonian MHD.
The simplifying assumption of the force-free limit has been applied 
as relativistic jets must be highly magnetized.
 
The central point of our paper is the consideration of
differential rotation of the foot points of the field lines,
i.e. a variation of the iso-rotation parameter $\omf(\Psi)$.
The underlying model is that of a magnetic jet anchored in an
accretion disk.
Two main problems had to be solved in order to calculate a
two-dimensional field distribution:
a) to determine the {\it a priori} unknown location of 
the light surface, 
b) the proper treatment of the regularity condition
along that light surface.
The light surface is the force-free equivalent of the Alfv\'en surface
and provides a singularity in the Grad-Shafranov equation.
We summarize our results as follows.

(1) We find numerical solutions for the two-dimensional magnetic flux 
distribution connecting the asymptotic cylindrical jet with a
differentially rotating disk.
In our example solutions the asymptotic jet radius is about 2.5 times
the asymptotic light cylinder radii.
This is the first truly two-dimensional relativistic solution for
a jet magnetosphere including differential rotation of the iso-rotation
parameter $\omf(\Psi)$.
The physical solution, being characterized by a smooth transition across the
light surface, is unique for a certain parameter choice for the rotation
law $\omf$.

(2) The half opening angle of the numerical jet solution is about 60
degrees. 
Cylindrical collimation is achieved already after a distance 
of 1-2 asymptotic light cylinder radii 
along the jet axis.
Differential rotation decreases the jet opening angle, but
increases the distance from the jet origin where collimation is 
achieved.
The ``jet expansion rate'', defined as the ratio of the asymptotic jet
radius to the jet radius at the jet origin, is about 10. 

(3) From the analytical treatment of the regularity condition
along the asymptotic branches of the light surface 
we derive a general estimate for the jet opening angle. 
We find that the jet half opening angle is larger than
$45\degr$ and increases 
for a steeper profile of the differential rotation $\omf$.

(4) Our two-dimensional ansatz, in combination with the treatment of 
differential rotation, allows for a connection of the asymptotic jet
solution with the accretion disk.
Certain disk properties can be deduced from the asymptotic jet
parameters. 
Examples are the disk toroidal magnetic field distribution, with a 
maximum at half of the disk radius
and the angular momentum flux per unit time and unit radius. 
This is interesting as a boundary condition for accretion disk models.
We find that most of the angular momentum is lost in the outer part
of the disk.

(5) Application of our model to the M87 jet gives good agreement
qualitatively.
From our numerical solution we derive an asymptotic light cylinder of
the M87 jet of about 50 Schwarzschild radii.
Collimation of the jet would be achieved after a distance of two
asymptotic light cylinder radii from the source.
This value is comparable with the observations, however, the
opening angle in our model is larger by a factor of two.

In summary, 
we have presented the first global two-dimensional solutions for a 
relativistic jet magnetosphere taking into account differential
rotation of the jet footpoints.
From our {\em jet} model we may determine certain physical quantities in 
the {\em disk} that are not possible to observe, as 
e.g. the angular momentum flux distribution at the disk-jet interface.
Comparison with the M87 jet shows that our model seems to be consistent
with the observations, therefore allowing for a derivation of the
collimation distance, the light cylinder radius and the jet expansion rate
for that example.
Clearly, such features as the time-dependent ejection of knots
and the interaction process between disk, jet and central source
cannot be answered by our approach.
Time-dependent relativistic MHD simulations of the whole 
``star''-disk-jet system would be necessary, however, such codes are
not yet fully developed.

\begin{acknowledgements}
Part of this work was initiated when C.F. was holding a guest stipend 
of the Sonderforschungsbereich (SFB) 328 of the University of Heidelberg.
E.M. acknowledges a grant (FE\,490/1-1) from the Deutsche
Forschungsgemeinschaft (DFG).
\end{acknowledgements}

\appendix

\section{numerical methods}
%
For the solution of the two-dimensional GS equation we apply the method
of finite elements as developed by Camenzind (1987) and 
Fendt et al. (1995).
Differential rotation $\omf(\Psi)$ implies two major complications for
the numerical computation. 
The first one is the fact that position and shape of the light surface
$D=0$ is not known {\it a priori}.
Along the light surface the boundary condition is the regularity
condition, which, however, itself depends on the two-dimensional 
solution $\Psi(x,z)$.
The second problem is the GS source term for the differential rotation,
containing the gradient of the magnetic flux, $|\nabla\Psi|^2$.
Compared to the case of rigid rotation, this introduces another
(and stronger) non-linearity in the GS equation.
Therefore, a fragile numerical convergence process can be expected.

An additional complication is that our grid of finite elements of
second order 
may be inadequate for a calculation of monotonous gradients
between the elements if the numerical resolution is too low.
However, for appropriate numerical parameters as grid size, element
size and iteration step size, we were finally able to overcome these
difficulties.

\subsection{Determination of the light surface}
%
Here we discuss the iteration procedure we use to determine the
location of the light surface.
Because the rotation law $\omf(\Psi)$ is prescribed, the
radius where the light surface, $D=0$,
intersects the jet boundary, $\Psi = 1$,
is known,
\begin{equation}
\xl(\Psi=1) = 1/\omf(\Psi=1).
\end{equation}
However, the corresponding position in $z$-direction is not known.
Some estimates can be made about shape and inclination of the
light surface in the limit of large radii (see Sect.\,2.4),
but a general solution is not yet known.

We start the iteration procedure calculating the inner solution
(defined as the field distribution inside the light surface)
with an outer grid boundary at $x=1$ (for comparison see Fig.\,1).
This choice is equivalent to the light cylinder in the case of rigid
rotation.
For differential rotation the radius $x=1$ is defined as asymptotic
light cylinder (for large $z$).
For low $z$-values the boundary $x=1$ is located inside the light
surface $\xl(\Psi) = 1/\omf(\Psi)$.
Along this outer boundary (of the {\em inner} solution),
we apply a homogeneous Neumann boundary condition. 
Usually, this implies that the field lines will cross that boundary
perpendicularly.
However, in our case the homogeneous
Neumann boundary condition transforms into the regularity condition 
{\em if} the boundary becomes equivalent to the singular light surface.
As shown in Fendt et al. (1995), this transformation applies
``automatically'' in our finite element code.
This is due to the facts that (i) finite element code solves the 
{\em integrated} GS equation
and (ii) the boundary integral, which is proportional 
to $D=1-x^2\omf^2$, vanishes along the light surface.

With the GS solution of the first iteration step we estimate the
deviation of the chosen outer boundary from the true
light surface by calculating $D=1-x^2\omf^2(\Psi)$.
For the lowest $z$-value prescribed, we know that $D=1-x^2\omf^2(\Psi=1)$.
Then, the outer grid boundary $(x,z)$ is slowly moved to a larger radius
with $\Delta x \sim D(x,z)^2$.
As a consequence of the different numerical grid, the field distribution
will change.  The value of $D$ will, however, decrease.
This procedure is repeated until $D$ is below a certain limit, $D\simeq0$.
Having obtained the solution inside the light surface, that field
distribution is taken as inner boundary condition for the outer solution.

\end{document}